\documentclass[
   ,final            
  ,numberedheadings 
  ]
  {aipproc}

\layoutstyle{6x9}

\usepackage{amssymb, amsmath}

\newcommand\pubnumber{IITB-PHY-TH-1401}
\newcommand\pubdate{\today}

\newcommand\pubblock{\rightline{\begin{tabular}{l} \pubnumber\\
         \pubdate  \end{tabular}}}

\usepackage{graphicx}

\newcommand{\vev}[1]{\langle #1 \rangle}

\newcommand{\be}{\begin{equation}}
\newcommand{\ee}{\end{equation}}
\newcommand{\bea}{\begin{eqnarray}}
\newcommand{\eea}{\end{eqnarray}}

\begin{document}
\pubblock

\title{Spontaneous parity breaking with broken supersymmetry : cosmological constraint\footnote{\textsl{Presented 
by UAY at  10th International Symposium 
on Cosmology and Particle Astrophysics (CosPA2013) Hawaii, and 
partly at Eleventh Conference on the Intersections of Particle and Nuclear Physics, (CIPANP 2012), Florida}}}

\classification{12.10.-g,12.60.Jv, 11.27.+d}
\keywords      {left-right symmetry, neutrino mass, domain walls, supersymmetry, metastable vacua}

\author{Urjit A. Yajnik}{
  address={Indian Institute of Technology Bombay, Mumbai 400076, India}
}

\author{Sasmita Mishra}{
  address={Indian Institute of Technology Bombay, Mumbai 400076, India},
  altaddress={Current address : Physical Research Laboratory, Ahmedabad 380009 India}
}

\author{Debasish Borah}{
  address={Indian Institute of Technology Bombay, Mumbai 400076, India},
  altaddress={Current address : Physics Department, Tezpur University, Tezpur 784028}
}

\begin{abstract}
Unified models incorporating the right handed neutrino in a symmetric
way generically possess parity symmetry. If this is broken spontaneously 
it results in the formation of domain walls in the early Universe, whose
persistence is unwanted. A generic mechanism for destabilisation of such
walls is a small pressure difference signalled by difference in the free energy
across the walls. It is interesting to explore the possibility of such effects
in conjunction with the effects that break supersymmetry in a phenomenologically
acceptable way. 
Realising this possibility in the context of several scenarios of supersymmetry
breaking results in an upper bound on the scale of spontaneous parity 
breaking, often much lower than the GUT scale. In the left-right symmetric 
models studied, the upper bound is  no higher than $10^{11}$GeV but a 
scale as low as $10^5$GeV is acceptable.
\end{abstract}

\def\thefootnote{\fnsymbol{footnote}}
\setcounter{footnote}{0}

\maketitle

\section{Left-right symmetry : a quick introduction}

Chirality seems to be an essential feature of fundamental physics,
thereby allowing dynamical generation of fermion masses. However 
the observed parity violation of the Standard Model (SM) is not warranted
by chirality. Discovery of neutrino masses in the past two decades 
strongly suggests the existence of right handed neutrino states.
The resulting parity balanced spectrum of fermions begs a parity 
symmetric theory and parity violation could then be explained to
be of dynamical origin. An interesting fact to emerge is that 
the see-saw mechanism generically  suggests an $M_R$ scale considerably 
smaller than the scale of  coupling constant unification in $SO(10)$. 
It is therefore appealing  to look for left-right symmetry as an 
intermediate stage in the
sequence of symmetry breaking, and explore the possible range of
masses acceptable for $M_R$. The crucial phenomenological question 
is, could the new symmetries  be within the accessible range of 
the LHC and the colliders of foreseeable future,
and hence deserve the name Just Beyond the Standard Model (JBSM)?


Left-right symmetric model\cite{Mohapatra:1980qe, Mohapatra:1979ia} needs a Supersymmetric extension
as an expedient for avoiding the hierarchy problem. The minimal set of Higgs superfields required, with
their $SU(3)\otimes SU(2)_L\otimes SU(2)_R\otimes U(1)_{\scriptscriptstyle{B-L}} $  is,
\begin{eqnarray}
  {  \Phi_i = ( 1, 2, 2, 0 ),} & \hspace{1em} & {  i
  = 1, 2,} \nonumber\\
  {  \Delta = ( 1, 3, 1, 2 ),} 
\hspace{1em} 
{   \bar{\Delta} = ( 1, 3, 1, - 2 ),} 
& \hspace{1em} &
 {  \Delta_c = ( 1, 1, 3, - 2 ),} 
\hspace{1em} 
  {  \bar{\Delta}_c = ( 1, 1, 3, 2 ),} \nonumber\\
\Omega = ( 1, 3, 1, 0 ),& \hspace{2em} &\Omega_c = ( 1, 1, 3, 0 )
\end{eqnarray}
and further details of the model can be found in the references.
Here we consider a model with the Higgs content which ensures spontaneous parity breaking, 
preserving electromagnetic charge invariance, and retaining $R$ parity \cite{Aulakh:1998nn}. 
It contains the two additional triplet Higgs fields introduced in the third line above.
We refer to this as ABMRS model.
Supersymmetric minima breaking $SU(2)_R$ symmetry are signaled by the ansatz
\begin{equation}
  \begin{array}{ccc}
    { \langle \Omega_c \rangle = \left( \begin{array}{cc}
      \omega_c & 0\\
      0 & - \omega_c
    \end{array} \right),} & \hspace{2em} {
    \langle \Delta_c \rangle = \left( \begin{array}{cc}
      0 & 0\\
      d_c & 0
    \end{array} \right),} & 
  \end{array}
\end{equation}
In this model, with an enhanced $R$ symmetry,  we are lead naturally to a see-saw 
relation    $M^2_{B - L} = M_{EW} M_R$.
This means Leptogenesis is postponed to a lower energy scale closer to
$M_{EW}$. Being generically below $10^9$ GeV, this avoids the gravitino mass bound 
but requires non-thermal leptogensis\cite{Sahu:2004sb}.

For comparison we also take an alternative model to this, considered in \cite{Babu:2008ep} 
where a  superfield $S(1,1,1,0)$ also singlet under parity is included
in addition to the minimal set of Higgs required. This is referred here as BM model. 

\section{Cosmology of breaking and Soft terms}
\label{sec:csmlgyBrk} 
SUSY breaking soft terms emerge below the SUSY breaking scale $M_S$.
We now proceed with the stipulation advanced in \cite{Yajnik:2006kc} that
the role of the hidden sector dynamics is not only to break SUSY but 
also break parity. This permits in principle a relation between observables
arising from the two apparently independent breaking effects.

The soft terms which arise in the two models ABMRS and BM may be
parameterized as follows 
%
\begin{equation}
\mathcal{L}_{soft}^1 
=~m_1^2 \textrm{Tr} (\Delta \Delta^{\dagger}) +
m_2^2 \textrm{Tr} (\bar{\Delta} \bar{\Delta}^{\dagger})
+ ~m_3^2 \textrm{Tr} (\Delta_c \Delta^{\dagger}_c) +
m_4^2 \textrm{Tr} (\bar{\Delta}_c \bar{\Delta}^{\dagger}_c)
\label{eq:eqnsoftone}
\end{equation}
\begin{equation}
\mathcal{L}_{soft}^2 
=\alpha_1 \textrm{Tr} (\Delta \Omega \Delta^{\dagger}) +
\alpha_2 \textrm{Tr} (\bar{\Delta} \Omega \bar{\Delta}^{\dagger})
+ ~\alpha_3 \textrm{Tr} (\Delta_c \Omega_c \Delta^{\dagger}_c) +
\alpha_4 \textrm{Tr} (\bar{\Delta}_c \Omega_c \bar{\Delta}^{\dagger}_c)
\label{eq:eqnsofttwo}\\
\end{equation}
\begin{equation}
\mathcal{L}_{soft}^3 
 ~\beta_1 \textrm{Tr} (\Omega \Omega^{\dagger}) +
\beta_2 \textrm{Tr} (\Omega_c \Omega^{\dagger}_c)
\label{eq:eqnsoftthree}
\end{equation}
\begin{equation}
\mathcal{L}_{soft}^4 
=S[\gamma_1 \textrm{Tr} (\Delta\Delta^{\dagger}) +
\gamma_2 \textrm{Tr} (\bar{\Delta}\bar{\Delta}^{\dagger})]
+ ~ S^*[\gamma_3 \textrm{Tr} (\Delta_c\Delta^{\dagger}_c) +
\gamma_4 \textrm{Tr} (\bar{\Delta}_c\bar{\Delta}^{\dagger}_c)]
\label{eq:eqnsoftfour}
\end{equation}
\begin{equation}
\mathcal{L}_{soft}^5 
= ~  {\tilde\sigma}^2 |S|^2
\label{eq:eqnsoftfive} 
\end{equation}
%
For ABMRS model the relevant soft terms are given by,
\begin{equation} 
\mathcal{L}_{soft} = \mathcal{L}_{soft}^1 + \mathcal{L}_{soft}^2
+ \mathcal{L}_{soft}^3
 \end{equation}  
For BM model the soft terms are given by,
\begin{equation} 
\mathcal{L}_{soft} = \mathcal{L}_{soft}^1  + \mathcal{L}_{soft}^4
+ \mathcal{L}_{soft}^5
 \end{equation}  
Using the requirement $\delta \rho \sim T_D^4$ we can constrain the 
differences between the soft terms in the Left and Right sectors 
\cite{Sarkar:2007ic, Sarkar:2007er}. In the BM model the
$S$ field does not acquire a vev in the physically relevant vacua
and hence the terms in eq.s (\ref{eq:eqnsoftfour}) and (\ref{eq:eqnsoftfive}) 
do not contribute to the vacuum energy. The terms in eq. (\ref{eq:eqnsofttwo})
are suppressed in magnitude relative to those in eq. (\ref{eq:eqnsoftthree})
due to having $\Omega$ vev's to one power lower. This argument assumes
that the magnitude of the coefficients $\alpha$ are such as to not mix
up the symmetry breaking scales of the $\Omega$'s and the $\Delta$'s.

To obtain orders of magnitude we have taken the $m_i^2$ parameters 
to be of the form $m_1^2 \sim m_2^2$
$\sim m^2$ and $m_3^2 \sim m_4^2$ $\sim m^{'2}$ \cite{Sarkar:2007er} 
with $T_D$ in the range $ 10 - 10^3$ GeV \cite{Kawasaki:2004rx}. 
For both the models we have taken the value of the $\Delta$ vev's as 
$d \sim 10^4$ GeV. For ABMRS model additionally we take $\omega \sim 10^6 $ GeV.
The resulting differences required for successful removal of domain walls
are shown in Table \ref{tab:DWalls}. 

\begin{table}
\begin{tabular}{p{.25\textwidth}c|c|c|cp{.3\textwidth}|cp{.3\textwidth}} 
\hline
$T_D/$GeV & $\sim$ & $10$ & $10^2$ & $10^3$ \\ \hline \hline
$(m^2 - m^{2\prime})/\mbox{GeV}^2$ & $\sim$ &
$10^{-4}  $  & $1$  & $10^{4} $\\ [1mm]
$(\beta_1 - \beta_2)/\mbox{GeV}^2 $ & $\sim$ &
$10^{-8}  $ & $10^{-4} $ & 
$1$ \\ [1mm] \hline \hline
\end{tabular} 
%
\caption{Differences in values of soft supersymmetry breaking parameters 
for a range of domain wall decay temperature values $T_D$. The 
differences signify the extent of parity breaking. 
}
\label{tab:DWalls}
\end{table}
We see from table \ref{tab:DWalls} that assuming both the mass-squared 
differences $m^2-m'^2$ and $\beta_1-\beta_2$ arise from the same dynamics,
$\Omega$ fields are the determinant of the cosmology. This is because
the lower bound on the wall disappearance temperature $T_D$ required
by $\Omega$ fields is higher and the corresponding $T_D$ is reached 
sooner.  This situation changes if for some reason $\Omega$'s do not
contribute to the pressure difference across the walls. The BM
model does not have $\Omega$'s and falls in this category. 

During the period of time in between destabilization of the DW and their
decay, leptogenesis occurs due to these unstable DW as discussed in
\cite{Cline:2002ia, Sarkar:2007er}. After the disappearance of the walls at
the scale $T_D$, electroweak symmetry breaks at a scale $M_{EW} \sim 10^2$ 
GeV and standard cosmology takes over.
In the next section we discuss the implementation of GMSB scenario 
for these models.

\section{Transitory domain walls}
Spontaneous parity breaking 
leads to formation of Domain walls which quickly dominate the energy density of 
the Universe. It is necessary for recovering standard cosmology that these walls
disappear at least before the Big Bang Nucleosynthesis (BBN). In an intrinsically 
parity symmetric theory difference in the vacua resulting in destabilisation is 
not permitted.
We may seek these effects to have arisen from the hidden sector and communicated
along with the messenger fields \cite{Mishra:2008be}. Constraints on the hidden 
sector model and the communication mechanism can be obtained in this way. 
Here we report on other possibilities.

There are several studies of wall evolution, and
an estimate of the temperature at which the walls may destabilise,
parametrically expressed in terms of the surface tension of the walls, in turn 
determined by the parity breaking scale $M_R$. By equating
the  terms leading to small symmetry breaking  discussed in the previous para with
this parametric dependence then gives a bound on $M_R$.


The dynamics of the walls in a radiation dominated universe is determined 
by two quantities : \cite{Kibble:1980mv},
\textit{Tension force} $f_T\sim \sigma/R $, where $\sigma$ is energy per 
unit area and $R$ is the average scale of radius of curvature, and
\textit{Friction force} $f_F \sim \beta T^4$ for walls moving with speed
$\beta$ in a medium of temperature $T$. 
The scaling law for 
the growth of the scale $R(t)$ on which the wall  complex is smoothed out,  
is taken to be \( R(t)\approx (G \sigma)^{1/2} t^{3/2} \).
Also, $f_F\sim 1/(Gt^2)$ and $f_T \sim ( \sigma /(G t^3))^{1/2}$. Then 
the pressure difference required to overcome the above forces 
and destabilise the walls is
\begin{equation}
\delta \rho_{\scriptscriptstyle{RD}} \ge G \sigma^2 \approx \frac{M_R^6}{M_{Pl}^2}
\sim M_R^4 \frac{M_R^2}{M_{Pl}^2}
\label{eq:eps-vsix}
\end{equation}

The case of matter dominated evolution is relevant to moduli fields 
copiously produced in generic string inspired models \cite{Kawasaki:2004rx} 
of the Universe.
A wall complex formed at temperature $T_i\sim M_R$ is assumed to have first relaxed 
to being one wall segment per horizon volume. It then becomes comparable in energy 
density  to the ambient matter density, due to the difference in evolution rates,
$1/a(t)$ for walls compared to $1/a^3(t)$ for matter. For simplicity also demand
that the epoch of equality of the two contributions is the epoch also of instability, 
so as to avoid dominance by domain walls. Thus we can set 
\(
M_{Pl}^{-2} T_D^4 \sim H^2_{eq} \sim \sigma^{\frac{3}{4}} H_i^{\frac{1}{4}} M_{Pl}^{-3}.
\)
The corresponding temperature permits the estimate of the required pressure difference, 
\begin{equation}
 \delta \rho_{\scriptscriptstyle{MD}} > M_R^4 \left(\frac{M_R}{M_{Pl}}\right)^{3/2}
\label{eq:eps-v11half}
\end{equation}
 Thus in this case we find $(M_R/M_{Pl})^{3/2}$ \cite{Mishra:2009mk}, a milder suppression
factor than in the radiation dominated case above.

\section{Parity breaking from Planck suppressed effects}
For a generic neutral scalar field $\phi$, the higher 
dimensional operators that may break parity have the simple form \cite{Rai:1992xw}
\( 
V_{eff} = \frac{C_5}{M_{Pl}} \phi^5 
\). 
But this is only instructional because in realistic
theories, the structure and effectiveness of such terms is conditioned by
Gauge invariance and supersymmetry and the presence of several scalar species.

One possibility is that the parity breaking operators arise at Planck scale
\cite{Mishra:2009mk}. We shall assume the structure of the symmetry breaking 
terms  as dictated by the Kahler potential formalism and treat the cases of 
two different kinds of domain wall evolution. Substituting the VEV's in the 
effective potential, we get
\begin{equation}
  V^R_{eff} \sim \frac{a(c_R + d_R)}{M_{Pl}}M_R^4M_W + \frac{a(a_R + d_R)}{M_{Pl}}
M_R^3M_W^2
\end{equation}
and likewise $R\leftrightarrow L$. Hence, with generic coefficients $\kappa$,
which for naturalness should remain order unity,
\begin{equation}
  \delta\rho \sim \kappa^A \frac{M_R^4M_W}{M_{Pl}}+{\kappa'}^A\frac{M_R^3M_W^2}{M_{Pl}}
\end{equation}

Then equating to $\delta \rho_{\scriptscriptstyle{RD}}$, $ \delta \rho_{\scriptscriptstyle{MD}}$
derived above,
\begin{equation}
 \kappa^A_{RD} > 10^{-10} \left( \frac{M_R}{10^6 {\rm GeV}}\right)^2
\end{equation}
For $M_R$ 
scale tuned to $~10^9$GeV  needed to avoid gravitino problem after 
reheating at the end of inflation, $\kappa_{RD}\sim 10^{-4}$, a 
reasonable constraint. but requires
$\kappa^A_{RD}$ to be $O(1)$ or unnaturally large for the scale of 
$M_R$ greater than the intermediate scale $10^{11}$GeV.

Next, 
\begin{equation}
 \kappa^A_{MD} > 10^{-2} \left( \frac{M_R}{10^6 {\rm GeV}}\right)^{3/2},
\end{equation}
which seems to be a modest requirement, but taking $M_R\sim10^9$GeV 
required to have thermal leptogenesis
without the undesirable gravitino  production, leads to unnatural 
$\kappa_{MD}>10^{5/2}$.

Concluding this section we note that the least restrictive requirement on
$\delta \rho $ is $\gtrsim (1\mathrm{MeV})^4 $ in order for the walls to not ruin BBN. 
This requirement gives a \emph{lower} bound on the 
$M_R$ scale, generically much closer to the TeV scale.

\section{Customized GMSB for Left-Right symmetric models} 
\label{sec:customgmsb}
The differences required between the soft terms of the Left and the
Right sector for the DW to disappear at a temperature $T_D$ 
are not unnaturally large. However the reasons for appearance of 
even a small asymmetry between the Left and the Right fields is hard to
explain since the original theory is parity symmetric. 
We now try to explain the origin of this small difference by
focusing on the hidden sector, and relating it to SUSY breaking.

For this purpose we assume that the strong dynamics responsible for SUSY
breaking also breaks parity, which is then transmitted to the visible sector
via the messenger sector and encoded in the soft supersymmetry breaking terms.
We implement this idea by introducing two singlet fields $X$ and $X'$, 
respectively even and odd under parity.
\begin{equation}
X \leftrightarrow X, \qquad X' \leftrightarrow -X'.
\end{equation} 
The messenger sector superpotential then contains terms
\begin{eqnarray}
W &=& \sum_n \left[ \lambda_n X \left( \Phi_{nL} \bar{\Phi}_{nL} 
+ \Phi_{nR} \bar{\Phi}_{nR}\right) \right.
\nonumber \\ &&
+ \left. ~\lambda'_n X' \left(\Phi_{nL} \bar{\Phi}_{nL} 
- \Phi_{nR} \bar{\Phi}_{nR} \right) \right]
\end{eqnarray} 
For simplicity, we consider $n=1$. The fields $\Phi_{L}$, $\bar\Phi_{L}$
and $\Phi_{R}$, $\bar\Phi_{R}$ are complete representations of
a simple gauge group embedding the L-R symmetry group. Further we
require that the fields labelled $L$ get exchanged with fields
labelled $R$ under an inner automorphism which exchanges
$SU(2)_L$ and $SU(2)_R$ charges, e.g. the charge conjugation operation 
in $SO(10)$. As a simple possibility we consider the case when
$\Phi_{L}$, $\bar\Phi_{L}$ (respectively, $\Phi_{R}$, $\bar\Phi_{R}$) 
are neutral  under $SU(2)_R$ ($SU(2)_L$). Generalization to other 
representations is straightforward.

As a result of the dynamical SUSY breaking we expect the fields
$X$ and $X'$ to develop nontrivial vev's and $F$ terms and hence
give rise to mass scales
\begin{equation}
\Lambda_X = \frac{\vev{F_X}}{\vev{X}},
\qquad
\Lambda_{X'} = \frac{\vev{F_{X'}}}{\vev{X'}}.
\end{equation} 
Both of these are related to the dynamical SUSY breaking scale $M_S$,
however their values are different unless additional reasons of symmetry
would force them to be identical. Assuming that they are different
but comparable in magnitude we can show that Left-Right
breaking can be achieved simultaneously with SUSY breaking being
communicated. 

In the proposed model, the messenger fermions receive respective mass 
contributions
\begin{eqnarray} 
m_{f_L} &=& |\lambda\vev{X} + \lambda^{\prime}\vev{X^{\prime}}|\\ \nonumber
m_{f_R} &=& |\lambda\vev{X} - \lambda^{\prime}\vev{X^{\prime}}|
 \end{eqnarray} 
while the messenger scalars develop the masses
\begin{eqnarray}
m_{\phi_L}^2 &=&  |\lambda\vev{X} + \lambda^{\prime}\vev{X^{\prime}}|^2
\pm  |\lambda\vev{F_X} + \lambda^{\prime}\vev{F_{X^{\prime}}}| \\ \nonumber
m_{\phi_R}^2 &=&  |\lambda\vev{X} - \lambda^{\prime}\vev{X^{\prime}}|^2
\pm  |\lambda\vev{F_X} - \lambda^{\prime}\vev{F_{X^{\prime}}}|
\end{eqnarray}
We thus have both SUSY and parity breaking communicated through these
particles.

As a result the mass contributions to the gauginos 
of $SU(2)_L$ and $SU(2)_R$ from  both the $X$  and $X'$ fields with 
their corresponding auxiliary parts take the simple form,
\begin{equation}
M_{a_{L}} = \frac{\alpha_a}{4 \pi} 
\frac{\lambda \langle F_X \rangle + \lambda^{\prime}
\langle F_{X^{\prime}} \rangle}{\lambda \langle X \rangle
+ \lambda^{\prime} \langle X^{\prime} \rangle}
\end{equation}
and
\begin{equation}
M_{a_{R}} = \frac{\alpha_a}{4 \pi}
\frac{\lambda \langle F_X \rangle - \lambda^{\prime}
\langle F_{X^{\prime}} \rangle}{\lambda \langle X \rangle
- \lambda^{\prime} \langle X^{\prime} \rangle} 
\end{equation}
upto terms suppressed by $\sim F/X^2$.
Here $a = 1, 2, 3$. 
In turn there is a modification to  
scalar masses, through two-loop corrections, expressed to leading
orders in the $x_L$ or $x_R$ respectively, by  the generic formulae
\begin{equation}
m^2_{\phi_{L}} =
2 \left( \frac{\lambda \langle F_X \rangle + \lambda^{\prime}
\langle F_{X^{\prime}} \rangle}{\lambda \langle X \rangle
+ \lambda^{\prime} \langle X^{\prime} \rangle}\right)^2
\left [ \left (\frac{\alpha_3}{4\pi}\right )^2 C_3^\phi +
\left (\frac{\alpha_2}{4 \pi}\right )^2
(C_{2L}^\phi)
 + \left (\frac{\alpha_1}{4 \pi}\right )^2 C_1^\phi \right ]
\label{eq:ModSclrMs1} 
\end{equation}

\begin{equation}
m^2_{\phi_{R}} =
2 \left( \frac{\lambda \langle F_X \rangle - \lambda^{\prime}
\langle F_{X^{\prime}} \rangle}{\lambda \langle X \rangle
- \lambda^{\prime} \langle X^{\prime} \rangle}\right)^2
\left [ \left (\frac{\alpha_3}{4\pi}\right )^2 C_3^\phi +
\left (\frac{\alpha_2}{4 \pi}\right )^2
(C_{2R}^\phi)
 + \left (\frac{\alpha_1}{4 \pi}\right )^2 C_1^\phi \right ]
\label{eq:ModSclrMs2} 
\end{equation}

The resulting difference 
between the mass squared of the left and right 
sectors are obtained as
\begin{equation}
\delta m_\Delta^2 
=
2 (\Lambda_X)^2 f(\gamma, \sigma)
\left\{  \left(\frac{\alpha_2}{4\pi}\right)^2
+ \frac{6}{5} \left( \frac{\alpha_1}{4 \pi} \right)^2 \right\}
\label{eq:Dm2Delta}
\end{equation}
where,
\begin{equation}
 f(\gamma, \sigma) = 
\left( \frac{ 1 + \textrm{tan}\gamma}{1+
\textrm{tan} \sigma}\right)^2
- \left( \frac{ 1 - \textrm{tan} \gamma}{ 1 -
 \textrm{tan} \sigma}\right)^2 
\label{eq:tanfnc}
\end{equation}
We have brought $\Lambda_X$ out as the representative mass scale
and parameterised the ratio of mass scales by introducing 
\begin{equation}
\textrm{tan}\gamma = \frac{\lambda^{\prime} 
\langle F_{X^{\prime}} \rangle}{\lambda \langle F_X \rangle},
\quad \textrm{tan} \sigma = \frac{\lambda^{\prime} 
\langle X^{\prime}\rangle}{\lambda \langle X \rangle}
\end{equation}
Similarly, 
\begin{equation}
\delta m_\Omega^2 = 2 (\Lambda_X)^2 
f(\gamma, \sigma)
\left( \frac{\alpha_2}{4\pi} \right)^2
\label{eq:Dm2Omega}
\end{equation}
In the models studied here, the ABMRS model will have contribution
from both the above kind of terms. The BM model will have contribution
only from the $\Delta$ fields.

The contribution to slepton masses is also obtained from eq.s (\ref{eq:ModSclrMs1})
and (\ref{eq:ModSclrMs2}).
This can be used to estimate the magnitude of the overall scale $\Lambda_X$
to be $\geq 30$ TeV \cite{Dubovsky:1999xc} from collider limits. 
Substituting this in the above formulae (\ref{eq:Dm2Delta}) and (\ref{eq:Dm2Omega})
we obtain the magnitude of the factor $f(\gamma, \sigma)$ required for
cosmology as estimated in table \ref{tab:DWalls}. The resulting values of 
$f(\gamma, \sigma)$ are tabulated in table \ref{tab:lambdasp}. We see that obtaining 
the values of $T_D$ low
compared to TeV scale requires considerable fine tuning of $f$. The natural 
range of temperature for the disappearance 
of domain walls therefore remains TeV or higher, i.e., upto a few order of magnitudes 
lower  than the  scale at which they form.

Consider for instance $T_D \sim 3\times 10^2$GeV, which allows $(m^2-m'^2)$
to range over $\sim 10^{2}$GeV$^2$ to $10^{3}$ GeV$^2$.
Consider two representative values of $\tan\gamma$ and $\tan\sigma$ for 
of $(m^2-m'^2)$. First, $(m^2-m'^2)=(2\pm 1.5)\times 10^{3}$GeV$^2$. This results in
sufficient paramour space for the $F$ and $X$ parameters. 
However when we consider 
$(m^2-m'^2)\sim 10$ GeV$^2$. We find that 
this requires the two parameters to be fine tuned to each other as $\tan \gamma \sim 0.4$ and $\tan \sigma > 3$. 
While this is specific to the particular scheme 
we have proposed for the communication of parity violation along with SUSY violation,
our scheme we believe is fairly generic and the results may persist for other implementations
of this idea.
\begin{table}
\begin{tabular}{cc|c|c|c} 
\hline
$T_D/$GeV & $\sim$ & $10$ & $10^2$ & $10^3$ \\ \hline \hline
Adequate $(m^2-m'^2)$ & $\phantom{\sim}$
& $ 10^{-7} $  & $ 10^{-3} $  & $10$\\ 
Adequate $(\beta_1-\beta_2)$  & $\phantom{\sim}$
& $10^{-11}$ & $ 10^{-7}$ & $10^{-3}$ \\ [1mm] \hline \hline
\end{tabular}
\caption{Entries in this table are the values of the parameter $f(\gamma,\sigma)$, 
required to ensure  wall disappearance at temperature $T_D$
displayed in the header row. The table
should be read in conjuction with table \ref{tab:DWalls},
with the rows corresponding to each other.}
\label{tab:lambdasp}
\end{table}

\section{Supersymmetry breaking in metastable vacua}
The dilemma of phenomenology with broken supersymmetry can be captured in the  
fate of $R$ symmetry generic to superpotentials \cite{Nelson:1993nf}. 
An unbroken $R$ symmetry in the theory is required for SUSY breaking. 
$R$ symmetry when spontaneously broken leads  to $R$-axions which are 
unacceptable. If we give up $R$ symmetry, the ground state remains supersymmetric.
The solution proposed in \cite{Nelson:1993nf, Intriligator:2007py}, is to 
break $R$ symmetry  mildly, governed by a small parameter $\epsilon$.
Supersymmetric vacuum persists, but this can be pushed far away in
field space.  SUSY breaking local minimum is ensured near the origin, 
since it persists in the limit $\epsilon\rightarrow 0$.
A specific example of this scenario \cite{Intriligator:2006dd}  referred to as ISS, 
envisages $SU(N_c)$ SQCD (UV free) with $N_f (>N_c)$ flavors such that 
it is dual to a $SU(N_f-N_c)$ gauge theory (IR free) so called
magnetic phase, with $N^2_f$
singlet mesons $M$ and $N_f$ flavors of quarks $q, \tilde{q}$.

Thus we consider a Left-Right symmetric model with ISS mechanism
as proposed in \cite{Haba:2011pr}. The particle content of the electric theory is
$ Q_L^a \sim (3,1,2, 1, 1), \quad \tilde{Q}^a_L \sim (3^*, 1, 2, 1,
-1) $ and 
$ Q_R^a \sim (1,3,1, 2, -1), \quad \tilde{Q}^a_R \sim (1,3^*,1, 2, 1)$.
where $a = 1, N_f$ with the gauge group $G_{33221}$. 
This SQCD has $N_c = 3$, and we need $N_f \geq 4$. 

For $N_f = 4$ the dual magnetic theory has Left Right gauge group 
$SU(2)_L \times SU(2)_R \times U(1)_{B-L}$ and the effective fields are the
squarks and nonet mesons carrying either the $SU(2)_L$ or the $SU(2)_R$
charges. The Left-Right symmetric renormalisable superpotential of this
magnetic theory is 
\begin{equation}
W^0_{LR} = h \mathrm{Tr} \phi_L \Phi_L \tilde{\phi}_L -h \mu^2 \mathrm{Tr}
\Phi_L+h \mathrm{Tr} \phi_R \Phi_R \tilde{\phi}_R -h \mu^2 \mathrm{Tr}
\Phi_R
\label{WLR1}
\end{equation}
After integrating out the right handed chiral fields, the
superpotential becomes
\begin{equation}
W^0_{L} = h \mathrm{Tr} \phi_L \Phi_L \tilde{\phi}_L -h \mu^2 \mathrm{Tr}
\Phi_L+ h^4 \Lambda^{-1} \mathrm{det}\Phi_R -h \mu^2 \mathrm{Tr} \Phi_R
\label{WL1}
\end{equation}
which gives rise to SUSY preserving vacua at 
\begin{equation}
\langle h \Phi_R \rangle = \Lambda_m \epsilon^{2/3} = \mu
\frac{1}{\epsilon^{1/3}}
\label{rightvev}
\end{equation}
where $\epsilon = \frac{\mu}{\Lambda_m}$.
Thus the right handed sector exists in a metastable SUSY breaking
vacuum whereas the left handed sector is in a SUSY preserving vacuum
breaking D-parity spontaneously.

We next consider \cite{Borah:2011aw} Planck scale suppressed terms that may signal parity breaking
\begin{equation} 
W^1_{LR} = f_L \frac{\mathrm{Tr}(\phi_L \Phi_L \tilde{\phi}_L) \mathrm{Tr}
\Phi_L}{\Lambda_m} +f_R \frac{\mathrm{Tr}(\phi_R \Phi_R \tilde{\phi}_R)
\mathrm{Tr} \Phi_R}{\Lambda_m}+ f'_L \frac{(\mathrm{Tr} \Phi_L
)^4}{\Lambda_m}+f'_R \frac{(\mathrm{Tr} \Phi_R )^4}{\Lambda_m}
\end{equation}
 The terms of order $\frac{1}{\Lambda_m}$ are given by
\begin{equation}
V^1_R =  \frac{h}{\Lambda_m} S_R [f_R(\phi^0_R
\tilde{\phi}^0_R)^2+f'_R\phi^0_R \tilde{\phi}^0_R S^2_R +(\delta^0_R
-S_R)^2((\phi^0_R)^2+(\tilde{\phi}^0_R)^2)]
\end{equation}

The minimization conditions give $\phi \tilde{\phi} = \mu^2 $ and $S^0
= -\delta^0$. Denoting $\langle \phi^0_R \rangle = \langle
\tilde{\phi}^0_R \rangle = \mu $ and $\langle \delta^0_R \rangle
=-\langle S^0_R \rangle = M_R$, we have 
\begin{equation}
V^1_R = \frac{hf_R}{\Lambda_m} (\lvert \mu \rvert^4 M_R +\lvert \mu
\rvert^2 M^3_R ) 
\end{equation}
where we have also assumed $f'_R \approx f_R$. For $ \lvert \mu \rvert
< M_R$ 
Thus the effective energy density difference between the two types of vacua is
\begin{equation}
\delta \rho \sim h(f_R-f_L) \frac{\lvert \mu \rvert^2
M^3_R}{\Lambda_m}
\end{equation}
Thus for walls disappearing in matter dominated era, we get
\begin{equation}
M_R < \lvert \mu \rvert^{5/9} M^{4/9}_{Pl} \sim 1.3 \times 10^{10}\; \mathrm{GeV}
\end{equation}
with $\mu\sim$Tev.
Similarly for the walls disappearing in radiation dominated era,
\begin{equation}
M_R < \lvert \mu \rvert^{10/21} M^{11/21}_{Pl} \sim 10^{11}\; \mathrm{GeV}
\end{equation}

\section{Conclusions}
We have pursued the possibility of left-right symmetric models as Just Beyond Standard Models 
(JBSM), not possessing a large hierarchy.  We also adopt the natural points of view that right 
handed neutrinos must be included in the JBSM in a symmetric way and that the required 
parity breaking to match low energy physics arises from spontaneous breakdown.
The latter scenario is often eschewed due to the domain walls it entails in the early 
Universe. We turn the question around to ask given that the domain walls occur,
what physics could be responsible for their successful removal without jeopardising
naturalness. 

We do not advance any preferred way to provide the small asymmetry required to
get rid of the domain walls. However it is interesting to correlate the possibility that
these small effects may be correlated to the supersymmetry breaking. We have considered
three models along these lines. One in which the hidden sector breaking of supersymmetry 
is at a low energy, and mediated by a gauge sector. Another in which the generic scale
of supersymmetry breaking is at Planck scale and the breaking effects are conveyed
purely through Planck scale suppressed terms. Finally we have also considered a
possible implementation of the scenarios in which the supersymmetry breaking is
not in a hidden sector but occurs due to a metastable vacuum protected from decay by
a large suppression of tunnelling.

The general message seems to be that the parity breaking scale in any case is not
warranted to be as high as required for a full unification in $SO(10)$ and further,
several scenarios suggest that left-right symmetry as the larger package incorporating
the SM may be within the reach of future colliders.

\begin{theacknowledgments}
A part of the reported work was carried out at IIT Gandhinagar. The research was partially 
supported by a  Department of Science and Technology grant.
\end{theacknowledgments}

\bibliographystyle{aipproc}

\begin{thebibliography}{20}
\expandafter\ifx\csname natexlab\endcsname\relax\def\natexlab#1{#1}\fi
\providecommand{\enquote}[1]{``#1''}
\expandafter\ifx\csname url\endcsname\relax
  \def\url#1{\texttt{#1}}\fi
\expandafter\ifx\csname urlprefix\endcsname\relax\def\urlprefix{URL }\fi
\providecommand{\eprint}[2][]{\url{#2}}

\bibitem[Mohapatra and Marshak(1980)]{Mohapatra:1980qe}
R.~N. Mohapatra, and R.~E. Marshak, \emph{Phys. Rev. Lett.} \textbf{44},
  1316--1319 (1980).

\bibitem[Mohapatra and Senjanovic(1980)]{Mohapatra:1979ia}
R.~N. Mohapatra, and G.~Senjanovic, \emph{Phys. Rev. Lett.} \textbf{44}, 912
  (1980).

\bibitem[Aulakh et~al.(1998)]{Aulakh:1998nn}
C.~S. Aulakh, A.~Melfo, and G.~Senjanovic, \emph{Phys. Rev.} \textbf{D57},
  4174--4178 (1998), \eprint{hep-ph/9707256}.

\bibitem[Sahu and Yajnik(2005)]{Sahu:2004sb}
N.~Sahu, and U.~A. Yajnik, \emph{Phys. Rev.} \textbf{D71}, 023507 (2005),
  \eprint{hep-ph/0410075}.

\bibitem[Babu and Mohapatra(2008)]{Babu:2008ep}
K.~S. Babu, and R.~N. Mohapatra, \emph{Phys. Lett.} \textbf{B668}, 404--409
  (2008), \eprint{0807.0481}.

\bibitem[Yajnik and Sarkar(2007)]{Yajnik:2006kc}
U.~A. Yajnik, and A.~Sarkar, \emph{AIP Conf. Proc.} \textbf{903}, 685--688
  (2007), \eprint{hep-ph/0610161}.

\bibitem[Sarkar and Yajnik(2007)]{Sarkar:2007ic}
A.~Sarkar, and U.~A. Yajnik, \emph{Phys. Rev.} \textbf{D76}, 025001 (2007),
  \eprint{hep-ph/0703142}.

\bibitem[Sarkar et~al.(2008)]{Sarkar:2007er}
A.~Sarkar, Abhishek, and U.~A. Yajnik, \emph{Nucl. Phys.} \textbf{B800},
  253--269 (2008), \eprint{0710.5410}.

\bibitem[Kawasaki and Takahashi(2005)]{Kawasaki:2004rx}
M.~Kawasaki, and F.~Takahashi, \emph{Phys. Lett.} \textbf{B618}, 1--6 (2005),
  \eprint{hep-ph/0410158}.

\bibitem[Cline et~al.(2002)]{Cline:2002ia}
J.~M. Cline, U.~A. Yajnik, S.~N. Nayak, and M.~Rabikumar, \emph{Phys. Rev. D}
  \textbf{66}, 065001 (2002).

\bibitem[Mishra et~al.(2009)]{Mishra:2008be}
S.~Mishra, U.~A. Yajnik, and A.~Sarkar, \emph{Phys.Rev.} \textbf{D79}, 065038
  (2009), \eprint{0812.0868}.

\bibitem[Kibble(1980)]{Kibble:1980mv}
T.~W.~B. Kibble, \emph{Phys. Rept.} \textbf{67}, 183 (1980).

\bibitem[Mishra and Yajnik(2010)]{Mishra:2009mk}
S.~Mishra, and U.~A. Yajnik, \emph{Phys. Rev.} \textbf{D81}, 045010 (2010),
  \eprint{0911.1578}.

\bibitem[Rai and Senjanovic(1994)]{Rai:1992xw}
B.~Rai, and G.~Senjanovic, \emph{Phys. Rev.} \textbf{D49}, 2729--2733 (1994),
  \eprint{hep-ph/9301240}.

\bibitem[Dubovsky et~al.(1999)]{Dubovsky:1999xc}
S.~L. Dubovsky, D.~S. Gorbunov, and S.~V. Troitsky, \emph{Phys. Usp.}
  \textbf{42}, 623--651 (1999), hep-ph/9905466.

\bibitem[Nelson and Seiberg(1994)]{Nelson:1993nf}
A.~E. Nelson, and N.~Seiberg, \emph{Nucl.Phys.} \textbf{B416}, 46--62 (1994),
  \eprint{hep-ph/9309299}.

\bibitem[Intriligator et~al.(2007)]{Intriligator:2007py}
K.~A. Intriligator, N.~Seiberg, and D.~Shih, \emph{JHEP} \textbf{0707}, 017
  (2007), \eprint{hep-th/0703281}.

\bibitem[Intriligator et~al.(2006)]{Intriligator:2006dd}
K.~A. Intriligator, N.~Seiberg, and D.~Shih, \emph{JHEP} \textbf{0604}, 021
  (2006), \eprint{hep-th/0602239}.

\bibitem[Haba and Ohki(2011)]{Haba:2011pr}
N.~Haba, and H.~Ohki, \emph{JHEP} \textbf{1108}, 021 (2011),
  \eprint{1104.5405}.

\bibitem[Borah and Yajnik(2011)]{Borah:2011aw}
D.~Borah, and U.~A. Yajnik, \emph{JHEP} \textbf{1112}, 072 (2011),
  \eprint{1107.5438}.

\end{thebibliography}

\end{document}